\documentclass[prl,twocolumn,superscriptaddress]{revtex4}

\usepackage{graphicx}
\usepackage{subfigure}

\begin{document}



\title{Evidence of $s$-wave superconductivity in ternary intermetallic La$_3$Pd$_4$Si$_4$}

\author{S. V. Taylor}
\altaffiliation[Also at: ]{Departamento de F\'{\i}sica, Facultad de Ciencias,
Universidad Central de Venezuela, Apartado 47586, Caracas 1041-A, Venezuela}

\affiliation{Centro de F\'{\i}sica, Instituto Venezolano de Investigaciones Cient\'{\i}ficas, Apartado 20632, Caracas
1020-A, Venezuela}

\author{J. F. Landaeta}
\author{D. Subero}
\author{P. Machado}
\affiliation{Centro de F\'{\i}sica, Instituto Venezolano de Investigaciones Cient\'{\i}ficas, Apartado 20632, Caracas
1020-A, Venezuela}

\author{E. Bauer}
\affiliation{Institute of Solid State Physics, Vienna University of Technology, A-1040 Wien, Austria}

\author{I. Bonalde}
\affiliation{Centro de F\'{\i}sica, Instituto Venezolano de Investigaciones Cient\'{\i}ficas, Apartado 20632, Caracas
1020-A, Venezuela}

\email[]{bonalde@ivic.gob.ve}

\date{3 may 2015}

\begin{abstract}
We measured the temperature dependence of the magnetic penetration depth of La$_3$Pd$_4$Si$_4$ down to $0.02 \, T_c$. We observe a temperature-independent behaviour below $0.25 \, T_c$,
which is a firm evidence for a nodeless superconducting gap in this material. The data display a very small anomaly around 1 K which we attribute to the possible presence of a superconducting impurity phase. The superfluid density is well described by a two-phase model, considering La$_3$Pd$_4$Si$_4$ and the impurity phase. The present analysis suggests that the superconducting energy gap of La$_3$Pd$_4$Si$_4$ is isotropic, as expected for conventional BCS superconductors.
\end{abstract}

\pacs{74.20.Rp, 74.25.Ha, 74.25.N-, 74.70.Xa}

\maketitle

\section{introduction}

Superconductivity in intermetallic compounds has become of great interest lately. Among the most relevant of these compounds are the iron (oxy)pnictides ($T_c$ up to 55 K), the quaternary borocarbides ($T_c$ up to 23 K) and binary MgB$_2$ ($T_c=39$ K). Other intermetallic superconductors of the types RET$_2$X$_2$ and RE$_3$T$_4$X$_4$ (RE=rare earth, T=transition metal, X=Si,Ge,Sn) have also received attention because of their magnetic and superconducting phases. In particular those crystallizing in orthorhombic U$_3$Ni$_4$Si$_4$ type of structure ($Immm$), of which the Ce-based systems have ground states showing Kondo effects and their nonmagnetic lanthanum counterparts have superconducting phases.

An example of isostructural intermetallic compounds are Ce$_3$Pd$_4$Si$_4$ and La$_3$Pd$_4$Si$_4$. The former shows quantum criticality due to Kondo effect and does not display superconductivity,  while the latter becomes superconductive below 2.15 K. Here we are concerned with the superconducting state of La$_3$Pd$_4$Si$_4$, which has been suggested to be described by a two-band model in a recent work on specific heat and critical fields \cite{bauer6}. This conclusion is consistent with recent electronic band-structure calculations that indicate multi-band Fermi surfaces in this material \cite{winiarski}. Two-gap superconductivity would be surprising in this material, since standard $s$-wave BCS superconducting states have been found in all other La-T-X compounds. Early DC magnetization and electrical resistivity measurements suggest that La$_3$Pd$_4$Si$_4$ is a type II superconductor \cite{fujii}.

In previous works it was suggested that impurity phases like LaPd$_2$Si$_2$ or La$_2$Pd$_3$Si$_3$ may exist in polycrystalline samples of La$_3$Pd$_4$Si$_4$. Although it was not possible to index some extra x-ray peaks to a LaPd$_2$Si$_2$ phase in Ref.~[\onlinecite{fujii}], the data clearly indicate the presence of an impurity phase. In another x-ray study  \cite{bauer6}, small extra peaks were indexed to La$_2$Pd$_3$Si$_3$. The impurity in La$_3$Pd$_4$Si$_4$ is not well established, but it seems sure that its presence is only of the order of a few percent or less.

Here, we intend to shed light on the energy gap structure of La$_3$Pd$_4$Si$_4$ by measuring the magnetic penetration depth of polycrystalline samples down to 0.02$T_c$. The magnetic penetration depth $\lambda(T)$ is a direct response of the Cooper pairs and is widely considered one of the most powerful probes for the superconducting energy gap structure. We found that the penetration depth of La$_3$Pd$_4$Si$_4$ may be explained by one-gap ($s$-wave) superconductivity in the presence of an impurity phase.

\section{Experimental Details}

The La$_3$Pd$_4$Si$_4$ sample ($T_c = 2.05$ K) measured in this work was the same used in Ref.~[\onlinecite{bauer6}]. Penetration depth measurements were carried out utilizing a 13.5 MHz tunnel diode oscillator \cite{mine7}. The magnitude of the AC probing field was estimated to be less than 5 mOe, and the DC field at the sample was reduced to around 1 mOe. The deviation of the penetration depth from the lowest measured temperature, $\Delta\lambda(T)=\lambda(T)-\lambda(T_{min})$, was obtained up to $T \sim 0.99T_c$ from the change in the measured resonance frequency $\Delta f(T)$: $\Delta f(T) = G\Delta \lambda(T)$. Here $G$ is a constant that depends on the sample and coil geometries and that includes the demagnetizing factor of the sample. To within this calibration factor, $\Delta \lambda(T)$ are raw data. We estimated $G$ by measuring a sample of known behaviour and of the same dimensions as the test sample.

\section{Results and Discussion}

\begin{figure}
\centering
\scalebox{0.55}{\includegraphics{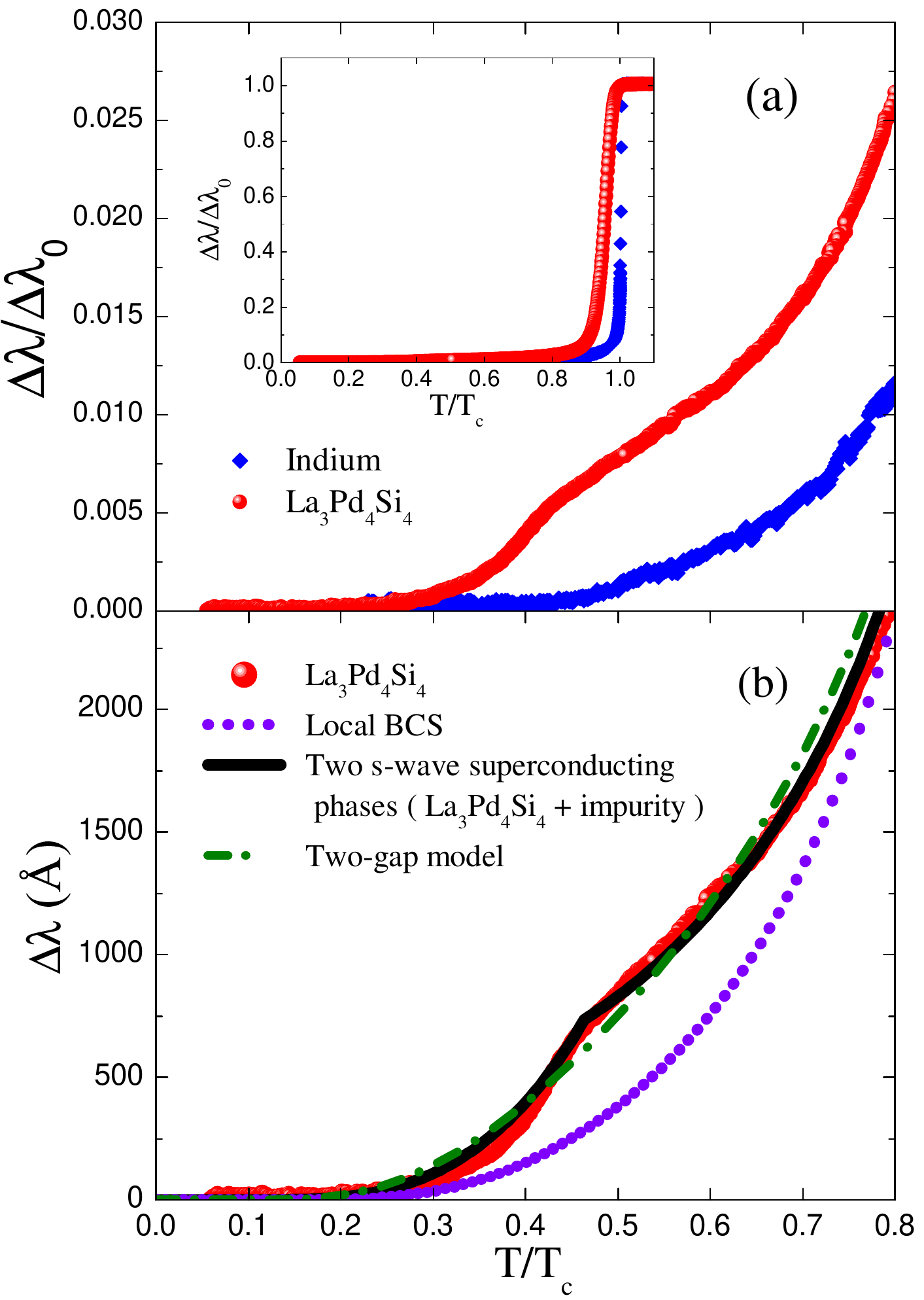}}
\caption{\label{figlambda}{(a) Magnetic penetration depth of In and La$_3$Pd$_4$Si$_4$ in the temperature region below 0.8$T_c$. The La$_3$Pd$_4$Si$_4$ data clearly expose an anomaly around 1 K. The inset displays the same sets of data in the whole temperature range below $T_c$.  (b) Penetration depth of polycrystalline La$_3$Pd$_4$Si$_4$ compared with numerical calculations of local BCS and two-gap models  and with simulation data of a model based on the existence of two superconducting phases (host + impurity) in the sample.}}
\end{figure}

The main panel of figure~\ref{figlambda}(a) shows the normalized deviation from its zero-temperature value of the penetration depth of La$_3$Pd$_4$Si$_4$ and pure indium (6N) polycrystalline samples below 0.8$T_c$. The inset of figure~\ref{figlambda}(a) depicts the same data in the entire superconducting region. A small (about 0.6$\%$ of the total signal) anomaly around 1 K is clearly observed in La$_3$Pd$_4$Si$_4$ sample. Since the indium data were taken under identical conditions and behave exactly as expected, the anomalous behaviour must comes from the La$_3$Pd$_4$Si$_4$ sample.

In Ref.~[\onlinecite{bauer6}] it was suggested that La$_3$Pd$_4$Si$_4$ may be described in terms of two-gap superconductivity. However, a two-gap model does not follow the low-temperature penetration depth data of this compound, as can be seen in figure~\ref{figlambda}(b). The experimental data are instead accounted for by a more general two-gap two-critical-temperature model, which in general suggest (a) a single phase with two independent superconducting energy bands or (b) two independent superconducting phases. The first scenario is an exotic situation never seen thus far, whereas the second one points out to the existence of a superconducting impurity in a sample. An example of the occurrence of the latter is Mo$_3$Al$_2$C \cite{mine19}. We believe scenario (b) is the most likely here in view of the proven existence of an impurity phase in La$_3$Pd$_4$Si$_4$ \cite{bauer6,fujii}. The small size of the anomaly is consistent with the very large superconducting volume assigned to the La$_3$Pd$_4$Si$_4$ phase \cite{fujii}.

Recent x-ray studies have discussed the possible presence of La$_2$Pd$_3$Si$_3$ \cite{bauer6} or LaPd$_2$Si$_2$ \cite{fujii} in La$_3$Pd$_4$Si$_4$ samples. To the best of our knowledge, in the La-Pd-Si system, apart from La$_3$Pd$_4$Si$_4$, only three other phases are superconductive: LaPdSi$_3$ ($T_c=2.6$ K) \cite{kitagawa}, LaPd$_2$Si$_2$ ($T_c=0.39$ K) \cite{palstra}, and La$_3$Pd$_5$Si ($T_c=1.4$ K) \cite{malik}. Interestingly, a careful analysis of the experimental data of La$_3$Pd$_5$Si in Ref.~[\onlinecite{malik}] indicates that in this compound both the zero resistivity and the peak of the specific heat jump occur around 1 K. These latter parameters are more closely related to the onset of the superconducting transition in susceptibility measurements. Hence, the anomaly around 1 K in the penetration depth of La$_3$Pd$_4$Si$_4$ may have been caused by small segregations of La$_3$Pd$_5$Si.

A re-analysis of the La$_3$Pd$_4$Si$_4$ x-ray spectrum did not confirm the presence of La$_3$Pd$_5$Si \cite{gribanovpc}.  However, the amount or volume fraction of the impurity in La$_3$Pd$_4$Si$_4$ is of the order of 1$\%$ or less, as detected here and in Ref.~[\onlinecite{fujii}], and could be too small to allow a detection by x-ray diffraction. Similarly, the resolution of the specific heat probe may not be enough to detect a tiny superconducting jump originated by a small amount of impurity, but be sufficient to observe a deviation from a pure s-wave behaviour.

The present analysis suggests that La$_3$Pd$_4$Si$_4$ is an $s$-wave superconductor. An important result here is that the penetration depth of La$_3$Pd$_4$Si$_4$ flattens out in the low-temperature region, below 0.2$T_c$ (main panel of figure~\ref{figlambda}(a)). This unambiguously indicates, regardless of the origin of the second drop, a \textit{nodeless} energy gap structure in La$_3$Pd$_4$Si$_4$. Moreover, it also implies that the superconducting impurity, say La$_3$Pd$_5$Si, has a nodeless energy gap too.

Next, we discuss our data in more detail by means of the superfluid density. We performed numerical calculations of the superfluid density for $s$-wave, two-gap, and two-phase models and compared them with the experimental data. For this purpose, we used the zero-temperature penetration depth $\lambda(0) = 378$ nm, as determined in Ref.~[\onlinecite{fujii}] and can be closely estimated from the London penetration depth $\lambda_L(0)=239$ nm and $l/\xi_0\approx0.75$ given in Ref.~[\onlinecite{bauer6}]. With a Ginzburg-Landau parameter $\kappa\approx 10$ \cite{bauer6}, La$_3$Pd$_4$Si$_4$ is a local superconductor. For such a superconductor the normalised superfluid density $\rho(T)= \left (n_s(T)/n \right )=\lambda^2(0)/\lambda^2(T)$, where $n$ is the total density, is given by

\begin{equation}
\label{supdens} \rho(T) = \sum_i N_i
\left [1 + 2\, \int^\infty_{\Delta_i}
\left ( \frac{\partial f_i}{\partial E_{i}} \right ) \frac{E_i}{(E^2_i - \Delta^2_i)^{1/2}} dE_i \right ] \,.
\end{equation}

\noindent Here, $N_i$ is the contribution to the total superfluid density of $i$ band (phase) and is related to the density of states on the corresponding Fermi surface. $f$ is the Fermi function. The total energy $E_i(T)=\sqrt{\epsilon^2_i + \Delta^2_i(T)}$, and
$\epsilon_i$ is the single-particle energy measured from the Fermi surface. We use here the standard weak-coupling gap interpolation formula $\Delta(T)=\Delta(0) \textrm{tanh}\left(\frac{\pi k_B T_c}{\Delta(0)} \sqrt{a(T_c/T - 1)}\right)$, where $\Delta(0)$ is the zero-temperature energy gap and $a$ is a constant related to the specific heat jump at the superconducting transition.

\begin{figure}
\centering
\scalebox{0.55}{\includegraphics{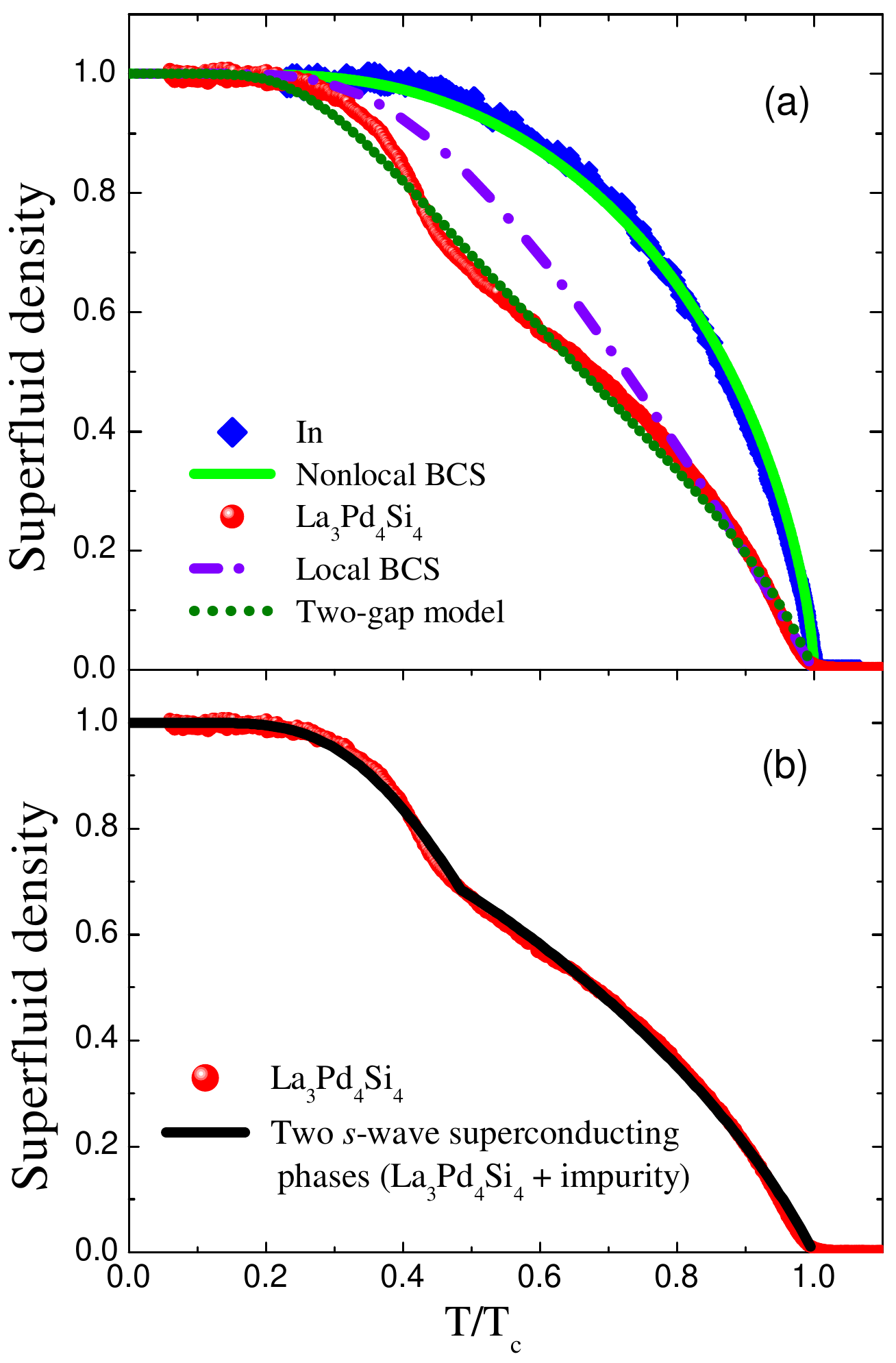}}
\caption{\label{fig:2}{Superfluid density of La$_3$Pd$_4$Si$_4$ compared with simulations of various models. (a) The In data and the numerical evaluation of a nonlocal BCS model are included for comparison. It is seen that local BCS and two-gap models cannot describe the La$_3$Pd$_4$Si$_4$ data. (b) A two-phase model based on the presence of two independent superconducting phases -host plus impurity- follows quite well the superfluid density of La$_3$Pd$_4$Si$_4$.}}
\end{figure}

For the local BCS model we used the standard BCS parameters. For the two-gap model we consider that the gaps are somehow coupled through interband interactions and the expression of $\Delta(T)$ given above differs for the two gaps only in the values of $\Delta(0)$. For this model Eq.~(\ref{supdens}) reduces for two bands $l$ and $s$ to $\rho=N\rho_l + (1-N)\rho_s$, where $N$ is the relative contribution of $\l$ band to the total superfluid density. For the two-phase model a second gap opens when the superconducting transition of the impurity phase takes place and in the expression for the gap interpolation formula, $\Delta(0)$, $a$ and $T_c$ are different for each phase.

Figure~\ref{fig:2} shows the superfluid density data and the results of the numerical calculations. For comparison, figure~\ref{fig:2}(a) also depicts the excellent agreement of the well established indium data and the simulation of the nonlocal BCS model (indium is a nonlocal superconductor), just to confirm the performance of the experimental system. Consistent with the penetration depth data, the superfluid density of La$_3$Pd$_4$Si$_4$ is temperature independent below 0.4 K. Also, the anomaly around 1 K in the penetration depth is displayed as a smooth bend in the superfluid density.

In figure~\ref{fig:2}(a) we compare the experimental data with the simulations of local BCS and two-gap models. The local BCS is in clear disagreement with the experimental data. For the best adjusting parameters the two-gap model does not follow the data quite well either, as opposed to what was suggested from the specific heat measurements \cite{bauer6}.

On the other hand, for the two-phase model the agreement with the data is remarkable in the whole temperature region, as seen in figure~\ref{fig:2}(b). In this case, $N=0.8$ is the normalized density of states of La$_3$Pd$_4$Si$_4$. The best values for $\Delta(0)/k_BT_c$ and $a$ are given in table~\ref{table}. Although in La$_3$Pd$_4$Si$_4$ the value of $\Delta(0)/k_BT_c$ is the one expected in the weak-coupling approximation, the parameter $a$ is much higher than expected in this approximation. The impurity phase would also be a conventional $s$-wave superconductor but in the strong-coupling limit, as indicated by the high values of $\Delta(0)/k_BT_c$ and $a$. In Ref.~[\onlinecite{malik}] La$_3$Pd$_5$Si was suggested to be a weak-coupling superconductor.

\begin{table}[t]
\caption{\label{table} Adjusting values of the model based on two superconducting phases.}
\vspace{5pt}
   \begin{center}
   \begin{tabular}{ccc}
   \hline
     & La$_3$Pd$_4$Si$_4$ & impurity  \\
     &       &   \\
     \hline
     $\Delta(0)/k_BT_c$ & 1.76 & 2.56 \\ [0.5ex]
    $a$ & 1.43 & 1.43  \\
    \end{tabular}
    \end{center}
\end{table}

Our results suggest that La$_3$Pd$_4$Si$_4$ is a conventional $s$-wave superconductors as all La-based intermetallic compounds, leaving exotic superconductivity to the Ce-based materials.

\section{Conclusions}

In summary, we performed magnetic penetration depth measurements in polycrystalline samples of La$_3$Pd$_4$Si$_4$. At the lowest temperatures of our measurements the superfluid density is temperature independent, indicating that the gap structure of La$_3$Pd$_4$Si$_4$ is nodeless. We found a second diamagnetic drop in the penetration depth that displays a smooth kink in the superfluid density. We suggest that the second drop maybe due to a superconducting impurity (possibly La$_3$Pd$_5$Si), and not necessarily a manifestation of two-band superconductivity.

\section{acknowledgments}
We appreciate conversations with C. Rojas and G. Gonz\'{a}lez. We also thank A. Argotte of the Electron Microscopy Facility at Instituto Venezolano de Investigaciones Cient\'{\i}ficas (IVIC). This work was supported by IVIC project no. 441 and by Austrian FWF grant no. P22295.


\begin{thebibliography}{9}
\expandafter\ifx\csname natexlab\endcsname\relax\def\natexlab#1{#1}\fi
\expandafter\ifx\csname bibnamefont\endcsname\relax
  \def\bibnamefont#1{#1}\fi
\expandafter\ifx\csname bibfnamefont\endcsname\relax
  \def\bibfnamefont#1{#1}\fi
\expandafter\ifx\csname citenamefont\endcsname\relax
  \def\citenamefont#1{#1}\fi
\expandafter\ifx\csname url\endcsname\relax
  \def\url#1{\texttt{#1}}\fi
\expandafter\ifx\csname urlprefix\endcsname\relax\def\urlprefix{URL }\fi
\providecommand{\bibinfo}[2]{#2}
\providecommand{\eprint}[2][]{\url{#2}}

\bibitem[{\citenamefont{Kneidinger et~al.}(2013)\citenamefont{Kneidinger,
  Michor, Bauer, Gribanov, Lipatov, Seropegin, Sereni, and Rogl}}]{bauer6}
\bibinfo{author}{\bibfnamefont{F.}~\bibnamefont{Kneidinger}},
  \bibinfo{author}{\bibfnamefont{H.}~\bibnamefont{Michor}},
  \bibinfo{author}{\bibfnamefont{E.}~\bibnamefont{Bauer}},
  \bibinfo{author}{\bibfnamefont{A.}~\bibnamefont{Gribanov}},
  \bibinfo{author}{\bibfnamefont{A.}~\bibnamefont{Lipatov}},
  \bibinfo{author}{\bibfnamefont{Y.}~\bibnamefont{Seropegin}},
  \bibinfo{author}{\bibfnamefont{J.}~\bibnamefont{Sereni}}, \bibnamefont{and}
  \bibinfo{author}{\bibfnamefont{P.}~\bibnamefont{Rogl}},
  \bibinfo{journal}{Phys. Rev. B} \textbf{\bibinfo{volume}{88}},
  \bibinfo{pages}{024423} (\bibinfo{year}{2013}).

\bibitem[{\citenamefont{Winiarski and Samsel-Czeka{\l}a}(2013)}]{winiarski}
\bibinfo{author}{\bibfnamefont{M.~J.} \bibnamefont{Winiarski}}
  \bibnamefont{and}
  \bibinfo{author}{\bibfnamefont{M.}~\bibnamefont{Samsel-Czeka{\l}a}},
  \bibinfo{journal}{J. Alloy. Compd.} \textbf{\bibinfo{volume}{546}},
  \bibinfo{pages}{124} (\bibinfo{year}{2013}).

\bibitem[{\citenamefont{Fujii}(2006)}]{fujii}
\bibinfo{author}{\bibfnamefont{H.}~\bibnamefont{Fujii}}, \bibinfo{journal}{J.
  Phys.: Condens. Matter} \textbf{\bibinfo{volume}{18}}, \bibinfo{pages}{8037}
  (\bibinfo{year}{2006}).

\bibitem[{\citenamefont{Bonalde et~al.}(2005)\citenamefont{Bonalde,
  Br\"{a}mer-Escamilla, and Bauer}}]{mine7}
\bibinfo{author}{\bibfnamefont{I.}~\bibnamefont{Bonalde}},
  \bibinfo{author}{\bibfnamefont{W.}~\bibnamefont{Br\"{a}mer-Escamilla}},
  \bibnamefont{and} \bibinfo{author}{\bibfnamefont{E.}~\bibnamefont{Bauer}},
  \bibinfo{journal}{Phys. Rev. Lett.} \textbf{\bibinfo{volume}{94}},
  \bibinfo{pages}{207002} (\bibinfo{year}{2005}).

\bibitem[{\citenamefont{Bonalde et~al.}(2011)\citenamefont{Bonalde, Kim,
  Prozorov, Rojas, Rogl, and Bauer}}]{mine19}
\bibinfo{author}{\bibfnamefont{I.}~\bibnamefont{Bonalde}},
  \bibinfo{author}{\bibfnamefont{H.}~\bibnamefont{Kim}},
  \bibinfo{author}{\bibfnamefont{R.}~\bibnamefont{Prozorov}},
  \bibinfo{author}{\bibfnamefont{C.}~\bibnamefont{Rojas}},
  \bibinfo{author}{\bibfnamefont{P.}~\bibnamefont{Rogl}}, \bibnamefont{and}
  \bibinfo{author}{\bibfnamefont{E.}~\bibnamefont{Bauer}},
  \bibinfo{journal}{Phys. Rev. B} \textbf{\bibinfo{volume}{84}},
  \bibinfo{pages}{134506} (\bibinfo{year}{2011}).

\bibitem[{\citenamefont{Kitagawa et~al.}(1997)\citenamefont{Kitagawa, Muro,
  Takeda, and Ishikawa}}]{kitagawa}
\bibinfo{author}{\bibfnamefont{J.}~\bibnamefont{Kitagawa}},
  \bibinfo{author}{\bibfnamefont{Y.}~\bibnamefont{Muro}},
  \bibinfo{author}{\bibfnamefont{N.}~\bibnamefont{Takeda}}, \bibnamefont{and}
  \bibinfo{author}{\bibfnamefont{M.}~\bibnamefont{Ishikawa}},
  \bibinfo{journal}{J. Phys. Soc. Jpn.} \textbf{\bibinfo{volume}{66}},
  \bibinfo{pages}{2163} (\bibinfo{year}{1997}).

\bibitem[{\citenamefont{Palstra et~al.}(1986)\citenamefont{Palstra, Lu,
  Menovsky, Nieuwenhuys, Kes, and Mydosh}}]{palstra}
\bibinfo{author}{\bibfnamefont{T.~T.~M.} \bibnamefont{Palstra}},
  \bibinfo{author}{\bibfnamefont{G.}~\bibnamefont{Lu}},
  \bibinfo{author}{\bibfnamefont{A.~A.} \bibnamefont{Menovsky}},
  \bibinfo{author}{\bibfnamefont{G.~J.} \bibnamefont{Nieuwenhuys}},
  \bibinfo{author}{\bibfnamefont{P.~H.} \bibnamefont{Kes}}, \bibnamefont{and}
  \bibinfo{author}{\bibfnamefont{J.~A.} \bibnamefont{Mydosh}},
  \bibinfo{journal}{Phys. Rev. B} \textbf{\bibinfo{volume}{34}},
  \bibinfo{pages}{4566} (\bibinfo{year}{1986}).

\bibitem[{\citenamefont{Malik and Kundaliya}(2003)}]{malik}
\bibinfo{author}{\bibfnamefont{S.~K.} \bibnamefont{Malik}} \bibnamefont{and}
  \bibinfo{author}{\bibfnamefont{D.~C.} \bibnamefont{Kundaliya}},
  \bibinfo{journal}{Solid State Commun.} \textbf{\bibinfo{volume}{127}},
  \bibinfo{pages}{279} (\bibinfo{year}{2003}).

\bibitem[{gri()}]{gribanovpc}
\bibinfo{note}{A. Gribanov, private communication}.

\end{thebibliography}

\end{document}